\documentclass[a4paper]{jpconf}
\usepackage{graphicx}
\usepackage{cite}
\begin{document}
\title{Trojan Horse method and radioactive ion beams: study of 
$^{18}$F(p,$\alpha$)$^{15}$O reaction at astrophysical energies}

\author{M. Gulino$^{1,2}$, S. Cherubini$^{1,3}$, G. G. Rapisarda$^{1,3}$, S. Kubono$^4$, L. Lamia$^3$, M. La Cognata$^1$, H. Yamaguchi$^5$, S. Hayakawa$^2$, Y. Wakabayashi$^5$, N. Iwasa$^6$, S. Kato$^7$, 
H. Komatsubara$^8$, T. Teranishi$^9$, A. Coc$^{10}$, N. De S\'er\'eville$^{10}$, F. Hammache$^{10}$, C. Spitaleri$^{1,3}$ 
}

\address{1) INFN-LNS, Catania, Italy\\
2) Universit\`a di Enna KORE, Enna, Italy\\
3) Dipartimento di Fisica e Astronomia, Universit\`a di Catania, Catania, Italy \\
4) RIKEN Nishina Center, 2-1 Hirosawa, Wako, Saitama 351-0198, Japan\\
5) Center for Nuclear Study (CNS), University of Tokyo, Wako Branch, 2-1 Hirosawa, Wako, Saitama 351-0198, Japan\\
6) Department of Physics, Tohoku University, 6-6 Aoba, Sendai, Miyagi 980-8578, Japan \\
7) Department of Physics, Yamagata University, Yamagata 990-8560, Japan \\
8) Institute of Physics, University of Tsukuba, Tennodai 1-1-1, Tsukuba, Ibaraki 305-8577, Japan\\
9) Department of Physics, Kyushu University, Fukuoka 812-8581, Japan \\
10) Centre de Spectrom\'etrie Nucl\'eaire et de Spectrom\'etrie de Masse, IN2P3, F-91405 Orsay, France
}
\ead{gulino@lns.infn.it}

\begin{abstract}
The Trojan 
Horse Method was applied for the first time to a Radioactive Ion Beam induced reaction to study 
the  reaction $^{18}$F(p,$\alpha$)$^{15}$O  via the 
three body reaction $^{18}$F(d,$\alpha$ $^{15}$O)n at the low energies relevant for astrophysics.\\
The abundance of $^{18}$F in Nova explosions is an important issue for the understanding 
of this astrophysical phenomenon. 
For this reason it is necessary to study the nuclear reactions that produce 
or destroy $^{18}$F in Novae.  $^{18}$F(p,$\alpha$)$^{15}$O
is one of the main $^{18}$F destruction channels. \\
Preliminary results are presented in this paper.
\end{abstract}

\section{Introduction}
The  $\gamma$-ray emission following the Nova explosion is dominated by the $511\, keV$ energy line, 
coming from the annihilation of positrons produced by the decay of radioactive nuclei. 
Among them, the $^{18}$F is especially important because of its expected abundance 
in the Nova environment and because of its lifetime, that well matches the timescale 
for Nova ejecta to become transparent to $\gamma$-ray emission. 
To understand the Nova explosion phenomena, it is then important 
to know the rate of nuclear reactions producing and destroying 
$^{18}$F. At relevant temperatures, the $^{18}$F(p,$\alpha$)$^{15}$O 
reaction is expected to dominate by roughly a factor of 1000 \cite{coc2000} and it is 
the uncertainty in this reaction rate that gives the main nuclear contribution to 
the overall uncertainty in the final abundance of $^{18}$F. 
The cross section of the reaction is dominated 
by the presence of several levels in the $^{19}$Ne compound nucleus
around the $^{18}$Fp threshold. Moreover, the tails 
 of some subthreshold levels are expected to contribute to the resonance cross section. 
 Especially important is not only the width of these resonance states but also their 
 interference effects. It is then important to explore the behavior of the cross section 
 in the center of mass energy range from $0$ up to $500\,keV$ to get information about 
 the reaction rate at Nova temperatures. 

In spite of several and long lasting attempts to measure the cross section of 
$^{18}$F(p,$\alpha$)$^{15}$O reaction (see e.g. 
\cite{wiescher1982,rehm1995,rehm1996,rehm1997,graulich1997,coszach1998, utku1998,
butt1998,graulich2001,bardayan2001,bardayan2002,kozub2005,kozub2005-2,desereville2003, 
beer2011,adekola2011}) the situation is still not satisfactory.
In order to get new complementary pieces of information 
on this process, we performed two experiments 
aiming at the measurement of the 
$^{18}$F(p,$\alpha$)$^{15}$O using two different 
techniques: the direct thick target method and the 
indirect Trojan Horse Method (THM). We report here 
only on the latter one. 
The THM was applied to a reaction induced by a Radioactive Ion Beam (RIB) for the first time 
in the experiment described here. Namely, the three body reaction $^{18}$F(d,$\alpha$ $^{15}$O)n
was used to infer information on the process of interest, $^{18}$F(p,$\alpha$)$^{15}$O,
in the low energy region important for astrophysics.

\section{The Trojan Horse Method}

\begin{figure}
\begin{center}
	\includegraphics[scale=0.3, trim=0.1cm 5.5cm 0.1cm 5.5cm, clip=true]{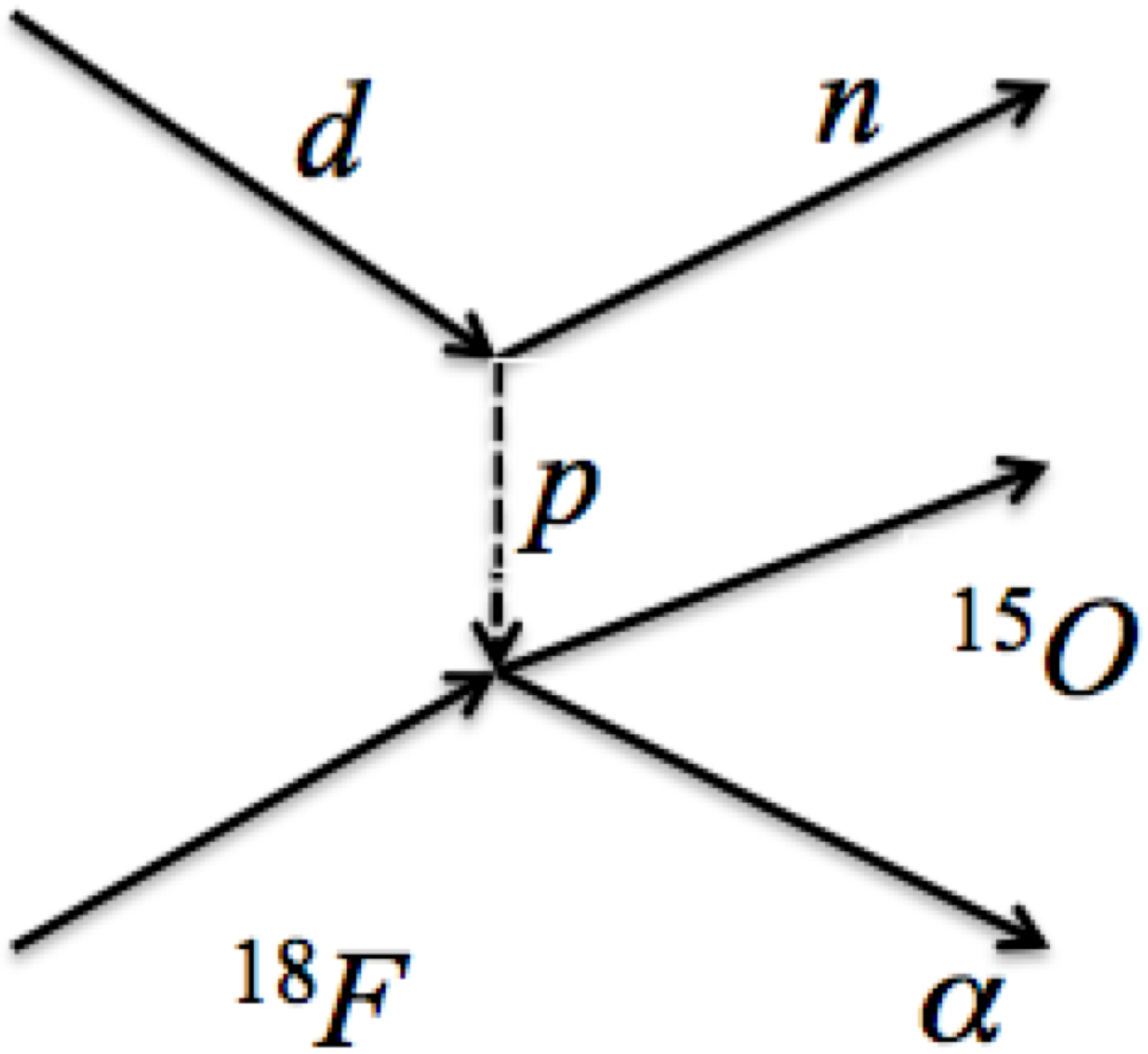}	
\end{center}
\caption{\label{thm} The THM basic diagram. The upper vertex describes the virtual decays
of particle $a$ into $x \otimes s$,  while the lower one refers to the process of interest, 
namely $b+x \rightarrow c+C$. In this paper, $a$ is deuteron going into $p$ and $n$ and the reaction 
of interest is $^{18}$F$+p \rightarrow ^{15}$O$+\alpha$.} 
\end{figure} 

The Trojan Horse Method (THM) \cite{Spitaleri_Folgaria90,Cherubini96,Baur_Typel04,Typel03,
Akram08} has 
been proposed and exploited to study the cross sections of nuclear reactions between 
charged particles 
at astrophysical 
energies. 
Briefly, a reaction of the type\\ 
\begin{equation}
x+b\rightarrow c +C  
\end{equation}
can be studied by using a 
three-body reaction of the type\\
\begin{equation}
  a+b\rightarrow c+C+s,
\end{equation}
where $a$ is a nucleus with a strong cluster structure of the type $x\otimes s$ and
it is often called the {\em trojan horse} nucleus.
The measurement of the reaction (2) is performed at energy higher than both the Coulomb barrier 
and the energy required to breakup particle $a$ into its components $x$ and $s$.
%

If reaction (2) proceeds through a quasi-free reaction mechanism, 
then it can be described 
by the diagram shown in figure \ref{thm} and
the cross section of the three-body reaction can be factorized in three parts:
a kinematical term, a term describing the virtual decay of $a$ into $x$ and $s$ and 
one giving the cross section of the reaction (1). 
Using a Plane Wave Impulse Approximation approach 
to describe the diagram in figure \ref{thm}, this can be written as \cite{Cherubini96,Baur_Typel04,Typel03}\\
\begin{equation}
 { d^3 \sigma \over{dE_c   d\Omega_c d\Omega_d} } \propto
 (KF) |\Phi(p_s) |^2  \left( {d \sigma \over{d\Omega_{cm}}} \right) ^{N}
\end{equation}
where $KF$ is the kinematical factor, $ |\Phi(p_s) |^2$ describes the impulse 
distribution of the spectator $s$ inside $a$ and $p_s$ is the momentum of the spectator. 
The symbol $N$ for the cross section of the reaction (1) indicates that 
it is a pure nuclear cross section, whitout effects coming from Coulomb interactions.
\begin{figure}[ht]
\begin{center}
{\includegraphics[scale=0.50, trim=4.1cm 1cm 4.1cm 1cm, clip=true, angle=270]{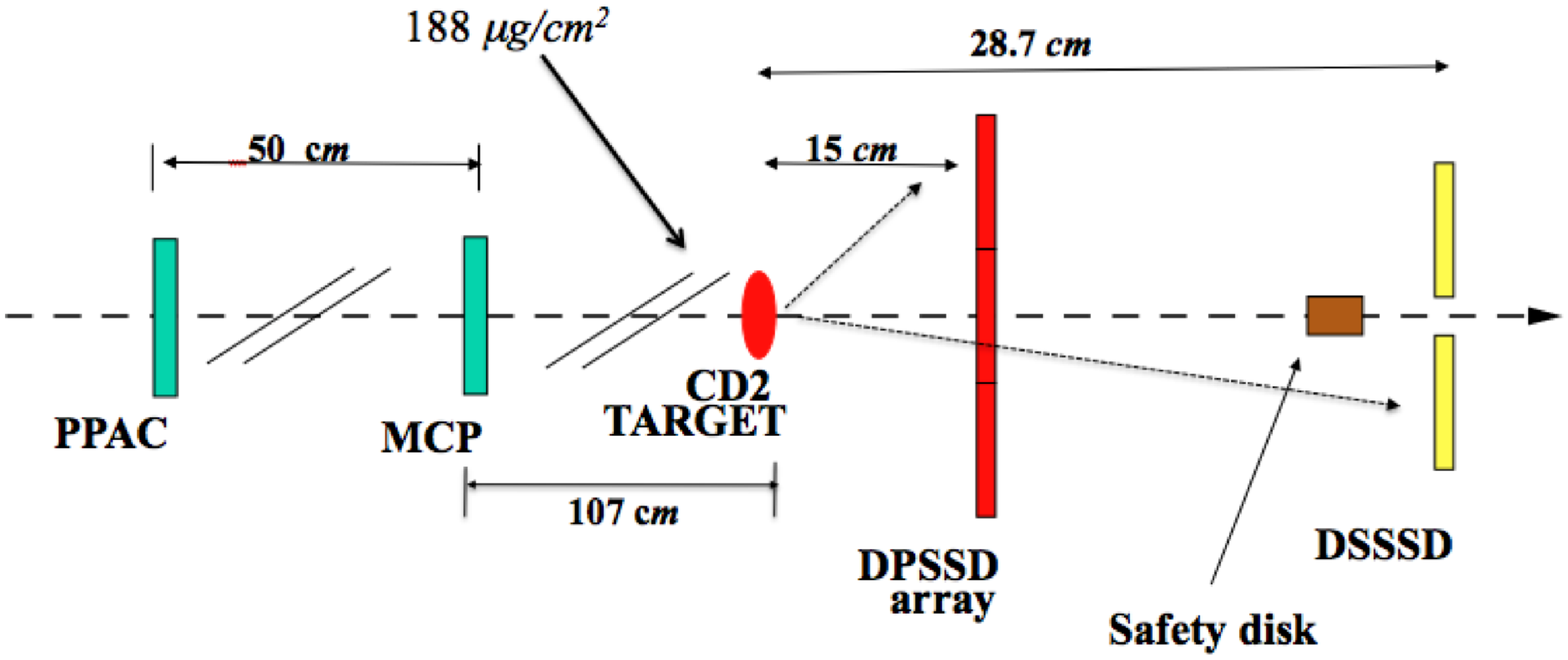}}
\end{center}
\caption{\label{astrho} Scheme of the experimental set-up used in the experiment.
The PPAC and MCP detectors were used for beam tracking,
hence allowing for
the reconstruction of the incidence angle of the beam on the CD$_2$ target (the
oval in the figure) 
 The reaction products are then detected by 
the plane made of bidimensional position sensitive detectors and by the two 
multistrip detectors. The safety disk protects the strips at very low angles from 
the risks that can derive from being hit by the direct beam.}
\end{figure}

Hence, by measuring the cross section of the three body process (2) and 
the impulse distribution of $s$ inside $a$ (represented by $|\Phi(p_s) |^2$), 
and by calculating the kinematical factor (KF), one can 
derive the cross section for the process of 
astrophysical interest (1). The c.m. interaction energy of the reaction (1) 
is given by the relation $E_{bx}=E_{cC}-Q_2$, where $ E_{cC}$ is the relative energy of
the outgoing particles $c$ and $C$ and $Q_2$ is the q-value of the two body process 
(1).
The calculated two body  cross section 
is a purely nuclear one, the 
Coulomb barrier having been overcome by
the TH mechanism \cite{Spitaleri_Folgaria90,Cherubini96,Baur_Typel04,
Akram08,Tumino07}. 
Hence, in order to compare the 
THM cross sections with the directly measured ones, the barrier suppression effects have to be
taken into account by correcting the THM cross section for the probability of penetration
under barrier.

%
 
 The THM has been used a number of times \cite{Cherubini96,Spitaleri99,Spitaleri01,
Lamia_Spitaleri04,Tumino07,Lacognata07,Lacognata08,Lacognata10, Spitaleri2011,Lamia12,ser10,cherubini11} 
to measure cross sections of 
astrophysical interest and 
it is presently regarded as a very powerful tool by the nuclear astrophysics community. 
Moreover the THM can be also used to study neutron induced reaction \cite{gul10}. 

The simple theoretical approach discussed above has been improved along the years.  
More sophisticated techniques have been introduced 
to treat more complicated situations as in the cases where the cross section is 
dominated by the presence of narrow, close and/or interfering resonances \cite{Lacognata08,Lacognata10}.

\section{The Experiment}

The  $^{18}$F beam was produced at the CRIB separator of the Center for
Nuclear Study (CNS) of the University of Tokyo installed at RIKEN campus in Wako,  
Japan, by using the 
$^{18}$O(p,n)$^{18}$F reaction. 
The $^{18}$O primary beam ($8^+$, E=4.5-5 $MeV/A$) was provided by a AVF cyclotron. It
impinged on a primary gas target of H$_2$. The reaction products were then
separated by the CRIB doubly achromatic spectrometer, described in detail in 
ref. \cite{yanagisawa2005}.
%

\begin{figure}[ht]
\begin{center}
{\includegraphics[scale=0.55, trim=6cm 0.1cm 6cm 0.1cm, clip=true, angle=270]{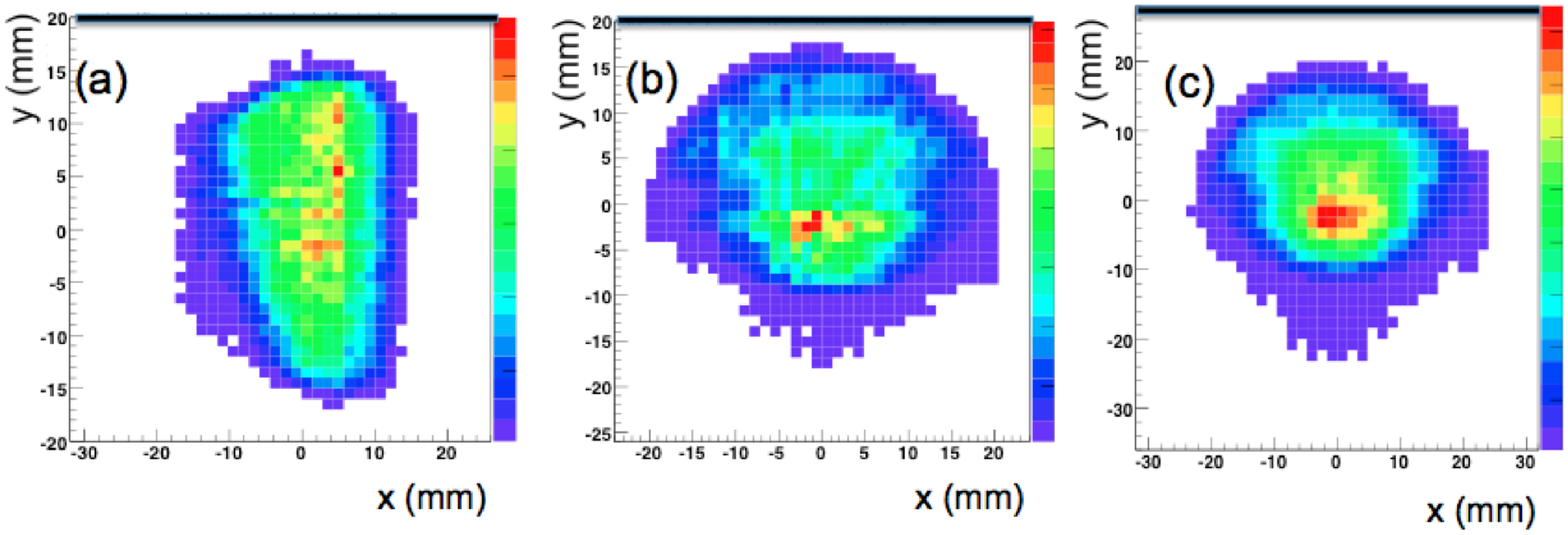}}
\end{center}
\caption{\label{beam}  Beam image on the PPAC detector (a), on the MCP detector (b) and 
beam spot on target position reconstructed event by event (c).}
\end{figure}
Two beam production tests were performed before the experiment. The characteristics of the 
 $^{18}$F beam 
used during the experiment were the following: beam energy peaked at $47.9\, MeV$ 
(FWHM $1.9 \,MeV$) with maximum beam intensity of
2$\times$10$^6$
pps and purity better than 98\%. The intensity was higher than 5$\times$10$^5$ pps throughout the 
measurement. This beam was used to bombard a CD$_2$ secondary target. 

The experimental set-up, schematically shown in figure (\ref{astrho}), 
 was very different from the typical ones used for THM experiments. 
Indeed, it consisted essentially of two parts: a beam 
tracker placed before the secondary target and two planes of position sensitive silicon detectors 
to detect the $^{15}$O and $\alpha$ ejectiles.

The tracks of the beam particles were reconstructed, event by event, using two Parallel Plate Avalanche 
Chambers (PPAC) with a position 
resolution ($x$-$y$) of roughly 1 $mm$ and a distance between the two planes of 500 $mm$. 
During the experiment the second PPAC broke down and it was replaced by a MicroChannel Plate 
detector (MCP) having the same position resolution. The MCP detected the electrons 
coming from the interaction of the beam with a Carbon foil placed in the same position of the broken PPAC. 
The precise particle tracking allowed for the determination of the 
intersection point of the projectiles on the secondary 
target. This is a key point for the application of the THM, as discussed 
in \cite{Omeg11}. 

The secondary target was made of a thin (roughly 150 $\mu$g/cm$^2$) CD$_2$ foil
of 30 mm diameter.


The detector setup was based on the modular 
system built at the INFN Laboratori Nazionali 
del Sud called ASTRHO (A Silicon Array for TRojan HOrse). In the configuration used for this 
experiment, ASTRHO consists of a mechanical frame that hosts two planes of position sensitive detectors.
The first plane consisted of 8 bidimentional position sensitive silicon detectors manufactured by Hamamatsu. Each detector is 
45$\times$45 mm$^2$ and measures the energy and the position on $x$ and $y$ directions of the 
impinging particle. The spatial resolution is of 1 $mm$ in both directions.  
The second plane was made of a pair of double sided silicon 
multistrip detectors (DSSSD), manufactured by Micron. Each detector (59$\times$50 mm$^2$) has 
16 $\times$16 strips in $x$ and $y$ direction respectively,
with a spatial resolution of each strip of 1.5 $mm$.

The detector holders were especially designed for this new apparatus, with the aim of having 
 a well defined geometry by construction and a support that can be adapted for different experiments by 
changing the distances among the different detector planes and the target.

In the present case the first plane of bidimensional position sensitive 
detectors was optimized to detect the $\alpha$ particles, while 
$^{15}$O is detected at most forward angles by DSSSD.
The angular ranges covered by the experimental set-up were $11^{\circ}<\theta<31^{\circ}$ for 
$\alpha$ particles and  $2^{\circ}<\theta<12^{\circ}$ for $^{15}$O ions.
For each detector the $x$ and $y$ position and the energy of the impinging particles were 
recorded together with the time of flight.

%
%

 \section{Data Analysis}
 
In order to apply the THM to the study of $^{18}$F(p,$\alpha$)$^{15}$O, 
we selected the $^{18}$F(d,$\alpha$ $^{15}$O)n three body reaction.

Given the low intensity of a RIB with respect to a stable beam, the use of a large solid angle
setup like ASTRHO was necessary to increase the global detection efficiency. 
Moreover the beam track reconstruction event by event was mandatory to reconstruct the 
angles of the ejected particles with respect to the actual incident particle trajectories, 
improving the angular resolution of the experiment.

In figure \ref{beam} the beam image reconstructed on the two detectors planes and on the target are reported. It is 
clearly seen that even if the beam spot on the target position is small, 
the beam divergence is not negligible.

\begin{figure}[ht]
\begin{center}
{\includegraphics[scale=0.3, trim=0.1cm 0.1cm 0.1cm 0.1cm, clip=true]{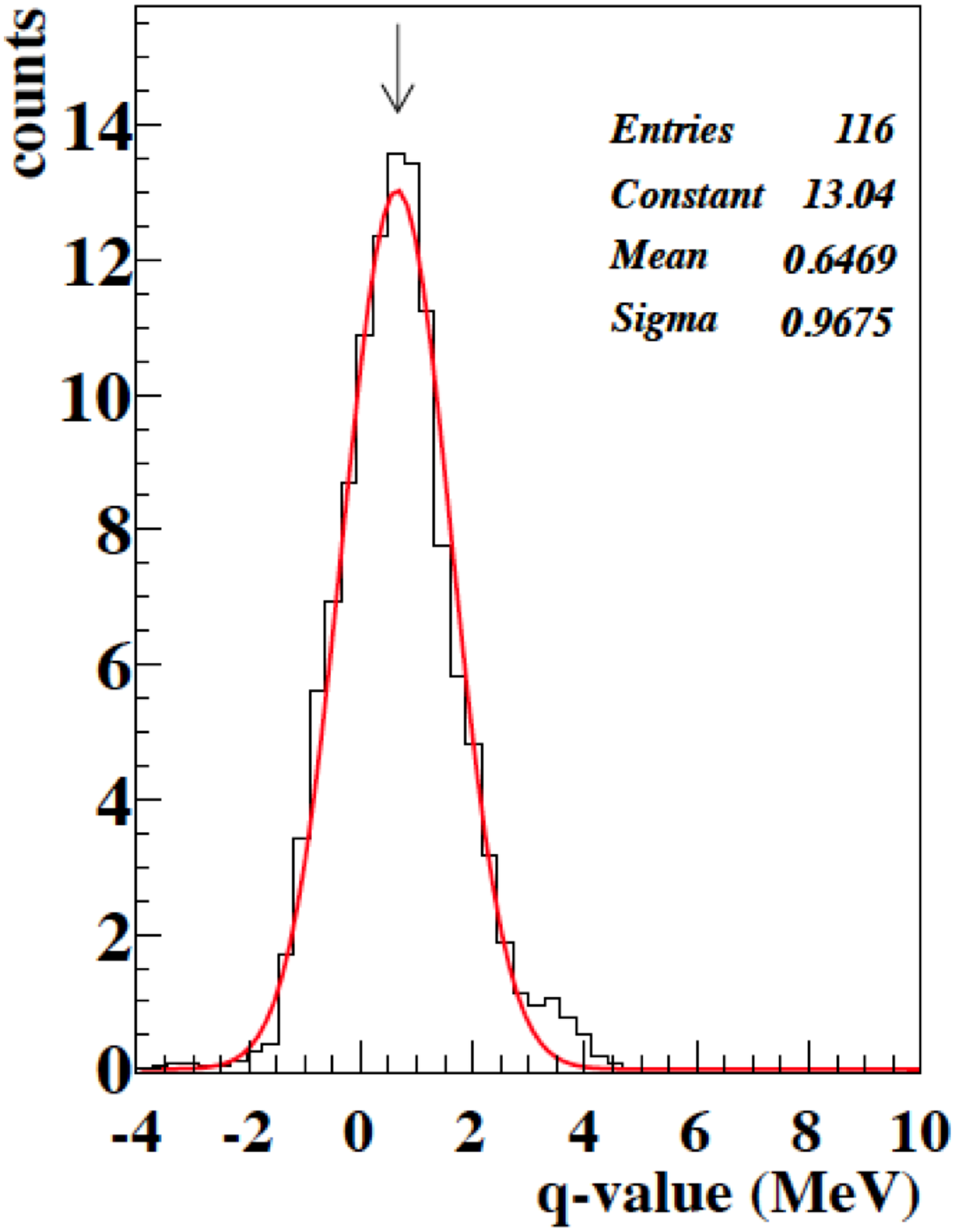}}
\end{center}
\caption{\label{qvalue} The Q-value spectra for the events that pass the conditions
imposed to identify the three body reaction channel of interest. The arrow represents the
position for the theoretical Q-value of the $^{18}$F(p,$\alpha$)$^{15}$O reaction
(0.658 MeV). The line is the result of a gaussian fit and the obtained parameters 
are reported in the figure.}
\end{figure}

The calibration in energy of the detectors was performed 
by using the elastic scattering peaks measured at different
beam energies for the reaction  $^{18}$O$+^{197}$Au.
 Moreover, a triple alpha source was used 
 to calibrate at low energies. The detectors made by Hamamatsu 
 were calibrated in position by using grids
placed in front of each detector during the calibration runs. 
The energy lost on the target by the beam, the
detected particle energy loss on the target and on the dead layer
of the detectors were also taken into account for the calibration in position
and energy.

After calibration, the data analysis of a THM experiment mainly proceeds 
through the following steps: identification of the process of interest, selection of the events 
corresponding to the quasi-free reaction channel, extraction of the two body cross section 
of astrophysical interest.

In the present experiment, due to the limitation in mass and Z identification of the 
outgoing particles, other three reactions with three bodies in the final state can 
give contribution in the phase space region where one expects events coming from
$^{18}$F(d,$\alpha$ $^{15}$O)n. These reactions are $^{18}$F(d,$\alpha$ $^{15}$N)p, 
$^{18}$F(d,p $^{18}$O)p and $^{18}$F(d,p $^{18}$F)n. 
We were able to select the channel of interest by applying convenient cuts in the phase space. 
Figure 4 shows the Q-value spectra that
was obtained by applying the event selection procedure. The position and width of the peak are
in very good agreement with the experimental conditions. The expected value for events
coming from  $^{18}$F(p,$\alpha$)$^{15}$O is 0.658 MeV
to be compared with a measured one of 0.647 MeV. Also, the measured width of the Q-value 
peak of FWHM 2.3 MeV is well
accounted for by the intrinsic beam energy spread (1.9 MeV FWHM) plus the kinematical 
enlargement. This gives us confidence of the correct identification of the events 
coming from the channel of 
interest.

\begin{figure}[ht]
\begin{center}
{\includegraphics[scale=0.34, trim=0.1cm 2cm 0.1cm 3cm, clip=true]{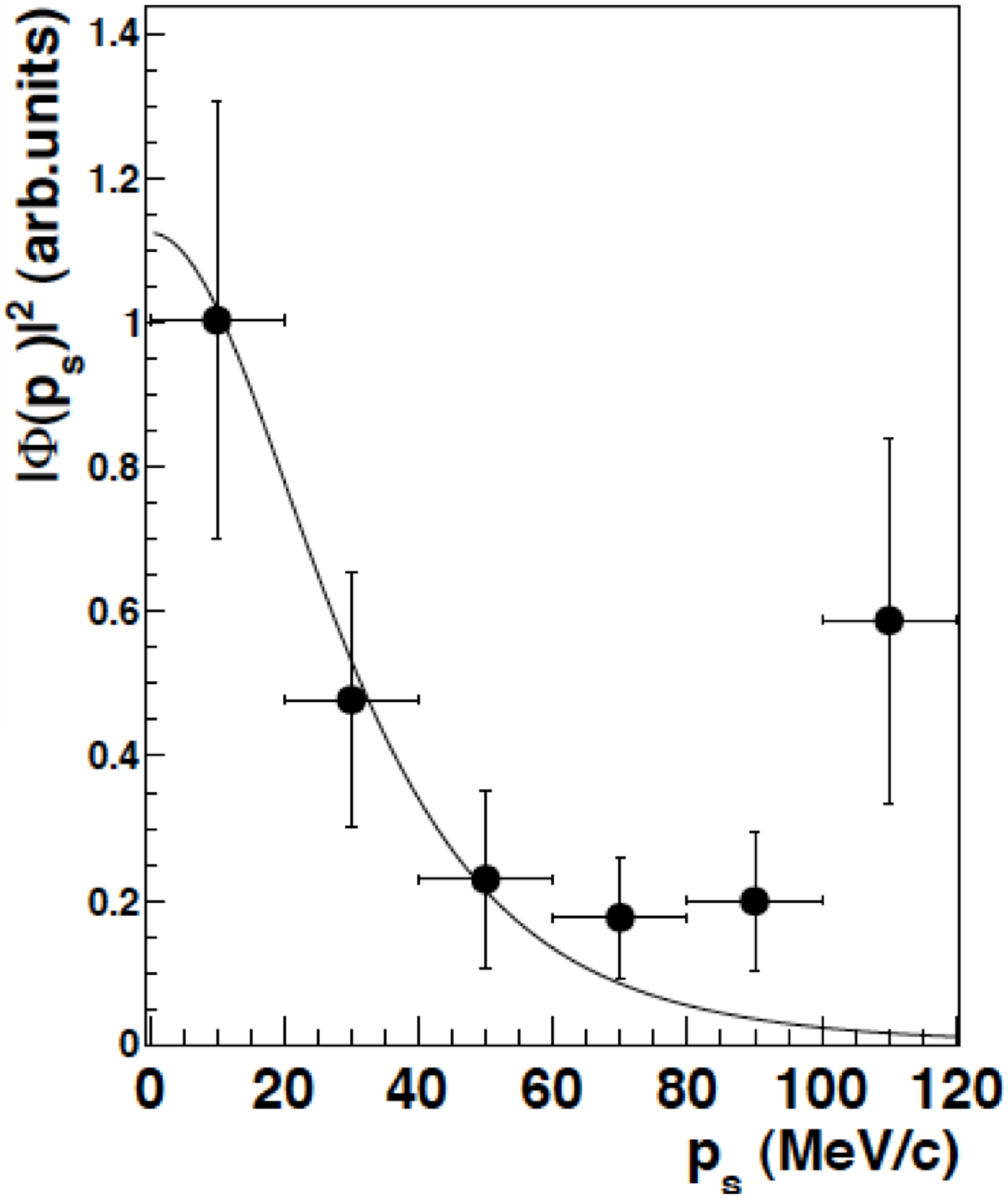}}
\end{center}
\caption{\label{ps} Momentum distribution for the $p$-$n$ intercluster
motion in deuteron. The solid line is the Hulth\'en function in momentum space. }
\end{figure}

The second step mentioned above in a THM analysis is the selection of events that
come from the quasi-free channel in the $^{18}$F(d,$\alpha$ $^{15}$O)n reaction,
among other processes falling the same particles in the final state.
The strongest evidence of the predominance of the 
 quasi-free mechanism 
is given by the shape of the momentum distribution for the $p$-$n$ intercluster
motion in deuteron. This is shown in figure \ref{ps} by black
dots as a function of the momentum of the spectator particle, $p_s$.

The solid line in figure \ref{ps} represents the
Hulth\'en function in momentum space with the standard parameters values \cite{zadro}. 
This is expected for deuteron as {\em trojan horse} nucleus 
if the plane-wave impulse approximation
is used to describe the reaction mechanism.  
Distortion effects influence the behavior of the distribution at high values of momentum, so
the rest of the analysis was done by imposing a cut on the momentum of the spectator
particle. Namely, only events where the spectator momentum was lower than 50 MeV/c 
were accepted.

\begin{figure}[ht]
\begin{center}
{\includegraphics[scale=0.35, trim=0.1cm 3.5cm 0.1cm 3.5cm, clip=true]{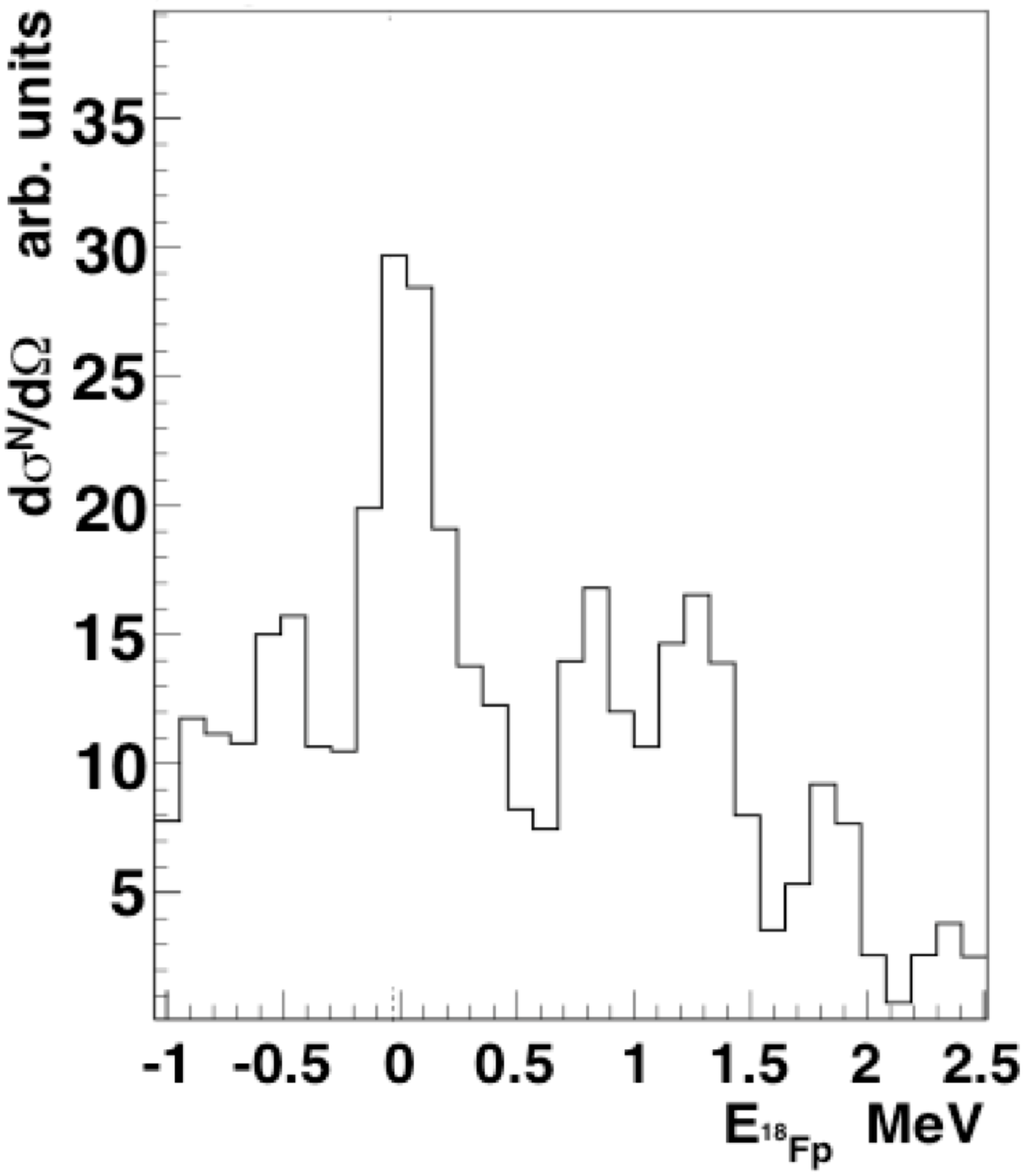}}
\end{center}
\caption{\label{xsec} The nuclear cross section spectrum in function of the p-$^{18}$F
cm energy for the events that pass 
the conditions described in the text.}
\end{figure}

Assuming the events selected according to the previous procedure are actually 
coming from the quasi-free contribution to the reaction yield, then one can apply
 equation (3) to obtain the cross section of interest for the 
$^{18}$F(p,$\alpha$)$^{15}$O process at astrophysical energies.
The result is shown in figure 6, that represents the nuclear cross section
for $^{18}$F(p,$\alpha$)$^{15}$O measured by THM down to zero (and even negative)
energies without any suppression due to the Coulomb barrier. 
In order to compare the present result with the direct data, it is necessary to correct the 
obtained nuclear cross section for the probability of penetration of the Coulomb and, 
possibly, centrifugal barrier. Data analysis is still in progress.

\section{Conclusion}
The THM was applied for the first time to study a reaction induced by a radioactive ion beam.
Even with the use of indirect methods like THM, the measurement of cross
sections of interest for nuclear astrophysics remains one of the most difficult tasks in nuclear
physics, especially if one has to use RIBs when the low beam intensity add on the top of 
the low cross sections typical of astrophysical nuclear processes. The use of a new large solid 
angle detector setup especially designed for this experiment was hence necessary 
in order to recover, at least in part, the possibility
of accumulating a statistics high enough to study the processes of interest.

The preliminary data analysis has demonstrated the feasibility of the experiment and 
gives us confidence that the results coming from the final analysis of the data will be helpful 
in understanding the problem of $^{18}$F destruction in Novae.

\section{Acknowledgments}
We are indebted to the AVF and CRIB technical staff for their invaluable contribution
to the experiment. Also we would like to thank the INFN-LNS workshop crew headed by Mr.B. Trovato
for the perfect construction of the ASTRHO mechanical frame and Mr. C. Marchetta from the target laboratory
at INFN-LNS for providing us with, among others, high quality and high purity CD$_2$ 
targets. The work was supported in part by the Italian
Ministry of University and Research under Grant No. RBFR082838 (FIRB2008).\\
\\


\begin{thebibliography}{99}
%
\bibitem{coc2000}A. Coc et al., A.\&A., 357 (2000) 561 and references
therein.
%
\bibitem{wiescher1982} M. Wiescher and K.U. Kettner, Astrophys. J. 263
(1982) 891.
%
\bibitem{rehm1995} K.E. Rehm, M. Paul, A.D. Roberts, et al., Phys. Rev.
C52 (1995) R460.
%
\bibitem{rehm1996} K.E. Rehm, M. Paul, A.D. Roberts, et al., Phys. Rev.
C53 (1996) 1950.
%
\bibitem{rehm1997} K.E. Rehm, C.L. Jiang, M. Paul, et al., Phys. Rev. C55
(1997) R566.
%
\bibitem{graulich1997} J.S. Graulich, F. Binon, W. Bradfield-Smith, et al.,
Nucl. Phys. A626 (1997) 751.
%
\bibitem{coszach1998} R. Coszach, M. Cogneau, C.R. Bain, et al., Phys.
Lett. B353 (1998) 184.
%
\bibitem{utku1998} S. Utku, J.G. Ross, N.P.T. Bateman, et al., Phys. Rev.
C57 (1998) 2731 and C58 (1998) 1354.
%
\bibitem{butt1998} Y.M. Butt, J.W. Hammer, M. Jaeger, et al., Phys. Rev.
C58 (1998) R10.
%
\bibitem{graulich2001} J.-S. Graulich, S. Cherubini, R. Coszach, et al.,Phys.
Rev. C63 (2001) 011302 and references therein.
%
\bibitem{bardayan2001} D.W. Bardayan, J.C. Blackmon, W. Bradfield-Smith,
et al., Phys. Rev. C63 (2001) 065802.
%
\bibitem{bardayan2002} D.W. Bardayan, J.C. Batchelder, J.C. Blackmon, et
al., Phys. Rev. Lett. 89 (2002) 262501.
%
\bibitem{desereville2003} N. De S\'ereville, A. Coc, C. Angulo, et al., Phys. Rev.
C67 (2003) 052801 (R) and references therein.
%
\bibitem{kozub2005} R.L. Kozub, D.W. Bardayan, J.C. Batchelder, et al.,
Phys. Rev. C71 (2005) 032801 (R).
%
\bibitem{kozub2005-2} R.L. Kozub, D.W. Bardayan, J.C. Batchelder, et al.,
Nucl. Phys. A 758 (2005) 753c-756c.
%
\bibitem{beer2011} C. E. Beer et al., Phys. Rev. C 83, 042801 (2011) 
\bibitem{adekola2011} A. S. Adekola, D. W. Bardayan, J.C. Blackmon, et al., Phys. Rev.
{\bf C83},  052801 (R), (2011).
%
%
\bibitem{Spitaleri_Folgaria90}
C.~Spitaleri.
 {\em in Proceedings of the Fifth Hadronic Physics Winter Seminar,
  Folgaria - TN, Italy}, Ed. World Scientific, Singapore, 1990.
%
\bibitem{Cherubini96}
S.~Cherubini, V.~N. Kondratyev, M.~Lattuada, C.~Spitaleri, D.~Miljanic,
  M.~Zadro, and G.~Baur.
 {\em ApJ}, {\bf 457}, 855, (1996)
%
\bibitem{Baur_Typel04}
G.~Baur and S.~Typel.
 {\em Prog. Theor. Phys. Suppl.}, {\bf 154}, 333, (2004)
%
\bibitem{Typel03}
 S.~Typel and G.~Baur.
\newblock {\em Ann. of Phys}, {\bf 305}, 228, (2003)
%
\bibitem{Akram08}
A. M. Mukhamedzhanov {\em et al.}
\newblock {\em Nucl. Phys.}, {\bf A787}, 321C, 2007.
%
\bibitem{Spitaleri99}
C.~Spitaleri, M.~Aliotta, S.~Cherubini, M.~Lattuada, Dj. Miljani\'c, S.~Romano,
  N.~Soi\'c, M.~Zadro, and R.~A. Zappal\`a, 
\newblock {\em Phys. Rev.} {\bf C60}, 055802, 1999.

\bibitem{Spitaleri01}
C.~Spitaleri {\em et al.},
\newblock {\em Phys. Rev.} {\bf C63}, 005801, 2001.

\bibitem{Lamia_Spitaleri04}
C.Spitaleri {\em et al.},
\newblock {\em Phys. Rev.} {\bf C69}, 055806, 2004.

\bibitem{Tumino07}
A.~Tumino et~al.,
\newblock {\em Phys. Rev. Lett.}, {\bf 98}, 252502, 2007.

\bibitem{Lacognata07}
M.~La~Cognata et~al.,
\newblock {\em Phys. Rev.} {\bf C76}, 06584, 2007.

\bibitem{Lacognata08}
M.~La~Cognata et~al.,
\newblock {\em Phys. Rev. Lett.}, {\bf 101}, 152501, 2008.

\bibitem{Lacognata10}
M.~La~Cognata et~al.,
\newblock {\em ApJ}, {\bf 708}, 796-811, 2010.

\bibitem{Spitaleri2011}

Adelberger, E.G., et al. {\em Solar fusion cross sections. II. the pp chain and CNO cycles}
(2011) Rev. Mod. Phys., 83 (1), pp. 195-245.


\bibitem{ser10} M.L.Sergi {\em et al.}, {\em Phys. Rev.} {\bf C82}, 032801 (2010)
\bibitem{Lamia12} L.Lamia {\em et al.},{\em  J. Phys. G: Nucl.Part.Phys.} {\bf 39}, 015106 (2012)
\bibitem{cherubini11} S. Cherubini {\em et al.},{\em Act. Phys. Pol. B} {\bf42},769 (2011)

\bibitem{gul10} M. Gulino {\em et al.},{\em J. Phys. G: Nucl.Part.Phys.} {\bf 37}, 125105 (2010)
\bibitem{yanagisawa2005} Y. Yanagisawa, S. Kubono, T. Teranishi, et al.,
Nucl. Instr. Meth. A 539 (2005) 74
\bibitem{Omeg11} S. Cherubini {\em et al.},{\em AIP Conf. Procee.} proceedings of OMEG11, in press (2012)
\bibitem{zadro} M. Zadro {\em et al.}, {\em Phys.Rev.} {\bf C40} 181, 1989
%
%
\end{thebibliography}
\end{document}